\documentclass[12pt]{article}
\usepackage{graphicx}

\begin{document}
\centerline {\large \bf Percolation and Galam Theory of Minority Opinion Spreading}

\bigskip
\centerline{Dietrich Stauffer}

\centerline{Institute for Theoretical Physics, Cologne University}

\centerline{D-50923 K\"oln, Euroland}

\bigskip
Abstract: The way in which an opinion rejecting reform can finally become the 
consensus of everyone was studied by Galam (2002) in a probabilistic model. 
We now replace his
clusters by those formed via random percolation by letting particles diffuse
on a lattice. Galam's  rejection of reform
is reproduced below the percolation threshold,
whereas at and above the threshold, also approval of a reform 
becomes possible.


\bigskip
In principle, the laws of a society should be based on the opinion of the 
majority. However, to protect minorities or to ensure stability, constitutions 
and other basic laws allow some changes only with a majority far above 50 
percent, and pose other hindrances. For example, in the USA the equal-rights
amendment to the constitution failed because not enough states did ratify it
in the alotted time. Galam [1,2], however, pointed out that even for normal
reforms which require only a majority of 50.01 percent for approval, this 
reform, due to the step-wise aspects of some opinion-forming and legislative
processes, is always blocked even though initially half of the people or more 
may be in favour. This blocking of change comes from the widespread simple rule 
that a tie vote means a ``no''. We confirm it here in a more realistic computer 
simulation. 

If in a democracy based on majority vote, a tie between yes and now means the
continuation of the status quo and the rejection of the reform, then in some
cases according to Galam [1,2] reform can become impossible even if initially
about half of the voters are in favour of reform. This happens [2] if the vote
proceeds in a hierarchical form, in which each group selects one representative
who represents at the next-higher level the majority opinion of the group, and
who votes for status quo if the represented group was evenly divided. In a less
artificial way, recently Galam [2] divided the population of $L$ individuals
into groups groups of various sizes $i$, with an arbitrary probability 
distribution function for $i$. Starting from a random distribution of opinions,
with 50 percent probability in favour,
each group determines its majority (biased by giving ``no'' in the case of 
a tie vote) and gives one common vote in an overall referendum. The resulting
fraction of group yes votes is the input for the probability of individuals to 
vote yes at the next time step. This procedure is iterated for several time
steps until a consensus is reached, and this consensus is always a ``no''. 
(See refs.3 for recent papers citing other consensus models, and refs.4 for
other examples of sociophysics.)

Now we replace the general and fixed group size distribution by the dynamical
one resulting from $x\,L^d$ individuals diffusing randomly on a $d$-dimensional 
hypercubic lattice of $L^d$ sites), and we take as groups the clusters of 
nearest neighbours forming in this way. Thus instead of a whole group size 
distribution we have just one concentration $x$ which determines the initial
cluster size distribution as in percolation theory. Initially, the individuals
favor ``yes'' with probability $p$ and no with probability $1-p$.
Thereafter, the diffusion 
process changes these clusters; in every time step, on average every particle
tries to move once to an empty neighbour site, and then the biased majority 
vote is taken. All individuals of each cluster adopt the majority opinion. This
adoption finishes this time step; afterwards the next time step starts with the
diffusion of individuals (each individual keeps its opinion when diffusing).

The bias against reform is strongest for small clusters where ties are frequent.
Large clusters favor yes and no with nearly equal probability. Thus we first
take the concentration $x$ such that the number of isolated pairs is maximal:
$x = 1/2d$. Then on the square lattice ($d=2, \, 23 \le L \le 3001, p=1/2$) we
found in all 20 samples that after about $10^2$ time steps everybody voted no.
Fig.1 shows the average time to reach consensus to increase as log$(L)$, perhaps
due to the fact that the largest cluster size below the percolation threshold 
increases as log$(L)$.

With instead $x$ equal to the percolation threshold 0.59, the results were 
mixed about half and half; the border between blocking of all reforms and
equal chance for blocking or success seems to be near $x=0.53$ (smaller for
small lattices).  At $x=1/4, \, L = 1001$ the initial opinion must be larger 
than 0.9999 in favor of reform to give it a chance.  
Also in higher dimensions at $x = 1/2d$ with $L=23, \, d=3,4,5$,
reform was always blocked.

\begin{figure}[hbt]
\begin{center}
\includegraphics[angle=-90,scale=0.5]{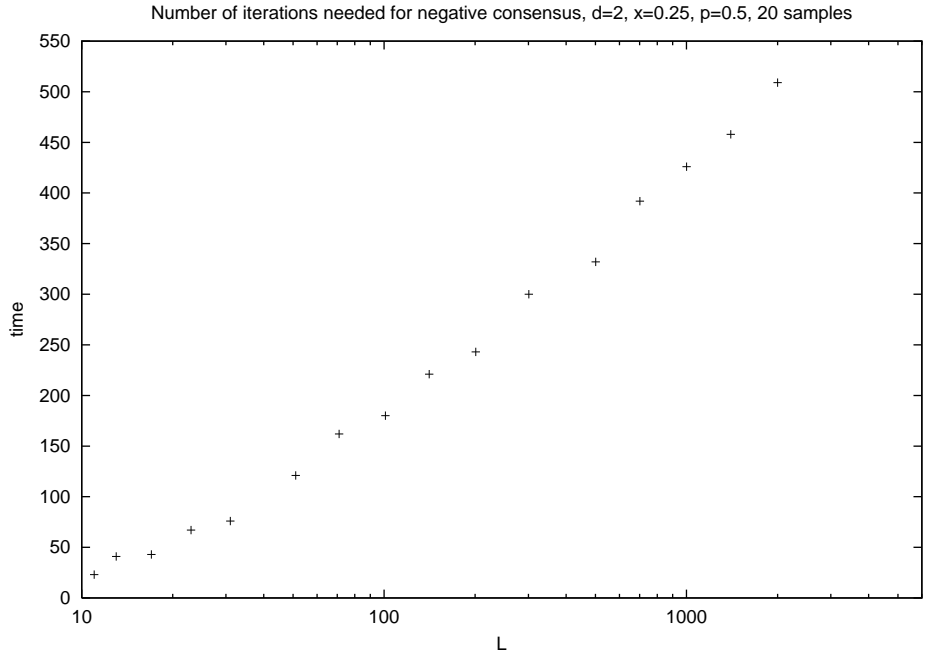}
\end{center}
\caption{
Logarithmic increase of the time needed to reach the negative consensus, versus
lattice size in two dimensions.
}
\end{figure}

In summary, we confirmed qualitatively the conclusion of Galam that reforms can 
be blocked by biased majority voting; but quantitatively our times to reach
this consensus are much longer than the examples given by Galam, and increase
logarithmically with system size. 

We thank Serge Galam for sending ref.1 before publication and for helpful 
remarks and suggestions. 
\bigskip
\parindent 0pt

[1] S. Galam, Eur. Phys. J. B 25, 403 (2002)

[2] S. Galam, J. Stat. Phys. 61, 943 (1990) and Le Monde, March 28, 2000,
pages 18-19; S. Galam and S. Wonczak, Eur. Phys. J. B 18, 183 (2000).

[3] G. Deffuant, D. Neau, F. Amblard and G. Weisbuch, Adv. Complex Syst. 3, 87
(2000); R. Hegselmann and M. Krause, Opinion dynamics and bounded confidence,
for Journal of Artificial Societies and Social Simulation 5 (2002); D. 
Stauffer, Int. J. Mod. Phys. C 13, issue 3 (2002).

[4] T.C. Schelling, J. Mathematical Sociology 1,143 (1971); J.M. Sakoda, 
ibidem 1, 119 (1971); S. Galam, Y. Gefen and Y. Shapir, ibidem 9, 1 (1982);
E. Callen and D. Shapero, Physics Today, July 1974, 23;
F. Schweitzer, (ed.) {\it  Self-Organization of Complex Structures:
From Individual to Collective Dynamics}, Gordon and Breach, Amsterdam 1997;
W. Weidlich, {\it Sociodynamics; A Systematic Approach to Mathematical Modelling
in the Social Sciences}. Harwood Academic Publishers, 2000

\end{document}